\documentclass[PAPER, atlasdraft=false, UKenglish, cernpreprint, texmf, orcidlogo]{atlasdoc}
\usepackage{atlaspackage}
\usepackage{atlasbiblatex}

\usepackage{atlasphysics}
\usepackage{multirow}
\newcolumntype{L}[1]{>{\raggedright\let\newline\\\arraybackslash\hspace{0pt}}m{#1}}
\newcolumntype{C}[1]{>{\centering\let\newline\\\arraybackslash\hspace{0pt}}m{#1}}
\newcolumntype{R}[1]{>{\raggedleft\let\newline\\\arraybackslash\hspace{0pt}}m{#1}}
\graphicspath{{logos/}{figures/}}

\usepackage{ANA-TAUP-2023-02-PAPER-defs}

\AtlasTitle{
Improved reconstruction of highly boosted $\tau$-lepton pairs in the $\tau\tau\rightarrow(\mu\nu_{\mu}\nu_{\tau})(\text{hadrons}+\nu_{\tau})$ decay channels with the ATLAS detector
}

\AtlasAbstract{
This paper presents a new $\tau$-lepton reconstruction
and identification procedure
at the ATLAS detector at the Large Hadron Collider,
which leads to significantly improved performance
in the case of physics processes where
a highly boosted pair of $\tau$-leptons is produced and one $\tau$-lepton decays
into a muon and two neutrinos ($\tau_\mathrm{\mu}$),
and the other decays into hadrons and one neutrino ($\tau_\mathrm{had}$).
By removing the muon information from the signals
used for reconstruction and identification of the
$\tau_\mathrm{had}$ candidate in the boosted pair, the efficiency
is raised to the level expected for an isolated $\tau_\mathrm{had}$.
The new procedure is validated by selecting
a sample of highly boosted
$Z\rightarrow\tau_\mathrm{\mu}\tau_\mathrm{had}$
candidates from the data sample of 140~$\mathrm{fb}^{-1}$
of proton--proton collisions at 13~TeV
recorded with the ATLAS detector.
Good agreement is
found between data and simulation predictions in both the
$Z\rightarrow\tau_\mathrm{\mu}\tau_\mathrm{had}$ signal region
and in a background validation region.
The results presented in this paper demonstrate the effectiveness
of the $\tau_\mathrm{had}$ reconstruction with muon removal in enhancing the signal sensitivity of the
boosted $\tau_\mathrm{\mu}\tau_\mathrm{had}$ channel at the ATLAS detector.
}

\AtlasRefCode{TAUP-2023-02}

\PreprintIdNumber{CERN-EP-2024-289}

\AtlasJournal{\EPJC}
\AtlasJournalRef{\EPJC 85 (2025) 706}
\AtlasDOI{10.1140/epjc/s10052-025-14012-4}

\AtlasCoverEgroupAnalysisTeam{atlas-taup-2023-02-analysis-team@cern.ch}

\hypersetup{pdftitle={ATLAS document},pdfauthor={The ATLAS Collaboration}}

\begin{document}

\maketitle

\tableofcontents

\newenvironment{DIFnomarkup}{}{}

\section{Introduction} \label{sec:intro}
The $\tau$-lepton is the heaviest known lepton,
with a mass of 1.777~$\GeV$, and a lifetime
of $2.9\times10^{-13}\text{ s}$~\cite{RevPartPhys}.
It has a 35\% probability of decaying leptonically
into a lighter lepton and two neutrinos:
$\tau\rightarrow e\nu_e\nu_\tau$ or
$\tau\rightarrow \mu\nu_\mu\nu_\tau$ (denoted $\taumu$),
collectively referred to as $\taulep$.
In the remaining 65\% of cases, the $\tau$-lepton
decays hadronically ($\tauhad$)
into one or more charged hadrons and zero or more neutral hadrons, plus one neutrino.
Thus, the fraction of $\tau$-lepton pairs resulting in a
$\tau_\mathrm{\mu}\tau_\mathrm{had}$ final state is 23\%.

The standard ATLAS $\tauhad$ reconstruction and identification
procedure~\cite{PERF-2013-06, ATLAS-CONF-2017-029, ATL-PHYS-PUB-2019-033, ATL-PHYS-PUB-2022-044}
starts with the reconstruction of the visible decay products of a
$\tauhad$ candidate (\tauhadvis), which is seeded by
a jet clustered using the anti-\(k_{t}\)
algorithm~\cite{Cacciari:2008gp, Fastjet} with a radius parameter $R=0.4$.\footnote{\AtlasCoordFootnote} The jet reconstruction algorithm operates on the
topological calorimeter cells~\cite{PERF-2014-07} calibrated to the local hadronic energy scale~\cite{LocalHadronicCali}.
This seed jet is referred to as $\tauseed$ jet. Only $\tauseed$ jets with transverse momentum $\pT > 5~\GeV$ and pseudorapidity $|\eta| < 2.5$ are considered.
Dedicated algorithms~\cite{ATL-PHYS-PUB-2022-044} are run to
identify the production vertex, associating and classifying reconstructed inner detector tracks,
and calibrating the energy of the $\tauhad$ candidate.
Tracks that have passed the $\tauseed$ jet association
requirements are classified into four categories: `Tau Tracks', `Conversion Tracks',
`Isolation Tracks' and `Fake Tracks'.
Only `Tau Tracks' are used to determine the number of charged decay products and thus the charge of the $\tau$ lepton.
After reconstruction, the $\tauhad$
candidate is classified by the
$\tauhad$ identification algorithm (TauID),
a recurrent neural network (RNN)
classifier~\cite{ATL-PHYS-PUB-2019-033},
to determine its likelihood of being
the decay product of a $\tauhad$.
Several working points,
`Tight', `Medium', `Loose' and `VeryLoose',
are then defined based on optimised requirements
on the RNN score with different $\tauhad$
identification efficiencies~\cite{ATL-PHYS-PUB-2019-033}.

The standard TauID algorithm is efficient unless activity from other particles is found inside the $\tauseed$ jet.
One of these cases is when a pair of $\tau$-leptons originates from a highly boosted resonance and the decay
products of the two $\tau$-leptons are reconstructed within the radius of a single $\tauseed$ jet. The
reconstruction and identification of boosted systems in which both $\tau$-leptons decay hadronically is
achieved by searching for hadronic $\tau\text{-like}$ substructure using a boosted decision tree within
a large radius seed jet~\cite{ATLAS-CONF-2020-012}. In this analysis, the decay into the $\tmth$ final state is considered.

Muon reconstruction~\cite{MUON-2018-03} in the ATLAS detector is based on the information from the inner tracking detector (ID) and the muon spectrometer (MS),
complemented by calorimeter data. The reconstruction begins by identifying tracks in the MS,
followed by matching these with tracks in the ID\@.
In cases where a candidate can be made from ID and MS tracks,
a combined fit is performed, taking into account the energy
loss in the calorimeters.
This study considers muons satisfying
the `Medium' identification working point~\cite{MUON-2018-03},
which ensures a combined muon where the ID track
matches the MS signature.
The minimum ionising nature of the muon and the fact that the muon reconstruction is independent of its isolation~\cite{MUON-2018-03}
enables the possibility of removing the ID track and clusters produced by the muon from the $\tauseed$ jet produced by a boosted $\tmth$ system. The
kinematic variables that are subsequently supplied to the TauID algorithm are re-calculated without the interference of
the muon. The $\tauhad$ reconstructed with this method is referred to as $\tauhadmurm$.

The $\tauhadmurm$ method was developed using simulated events corresponding to a beyond the Standard Model (BSM) process with a
high-mass graviton~\cite{Graviton_theory} decaying into two Higgs bosons as signal. Before using this technique in
future BSM searches, it is important to demonstrate its performance using a Standard Model (SM) process. As a validation
for the new method, the production of two $\tau$-leptons
originating from a highly boosted $Z$ boson
giving the final state of interest, $\Zttmuhad$, is used.

This paper is organised as follows.
The ATLAS detector is described briefly in Section~\ref{sec:detector}. The data and simulated samples are described in Section~\ref{sec:DataAndMCs}.
The development of the boosted $\tauhadmurm$ method is described in Section~\ref{sec:murm}.
The analysis methods and the results for the $\Zttmuhad$ validation are described in Section~\ref{sec:Zttbench}.
Finally, the conclusions are given in Section~\ref{sec:conclusion}.

\section{ATLAS detector} \label{sec:detector}

The ATLAS detector~\cite{PERF-2007-01}
at the Large Hadron Collider
(LHC)~\cite{Evans:2008zzb}
is a multipurpose particle detector
with a forward--backward symmetric
cylindrical geometry and a near \(4\pi\)
coverage in solid angle.
It consists of an ID surrounded by a thin superconducting solenoid
providing a \qty{2}{\tesla} axial magnetic field, electromagnetic (EM) and hadronic calorimeters, and a MS.
The inner tracking detector covers the pseudorapidity range \(|\eta| < 2.5\).
It consists of silicon pixel, silicon microstrip, and transition radiation tracking detectors.
Lead/liquid-argon (LAr) sampling calorimeters provide EM energy measurements
with high granularity within the region \(|\eta|< 3.2\).
A steel/scintillator-tile hadronic calorimeter covers the central pseudorapidity range (\(|\eta| < 1.7\)).
The endcap and forward regions are instrumented with LAr calorimeters
for EM and hadronic energy measurements up to \(|\eta| = 4.9\).
The MS surrounds the calorimeters and is based on
three large superconducting air-core toroidal magnets with eight coils each.
The field integral of the toroids ranges between \num{2.0} and \qty{6.0}{\tesla\metre}
across most of the detector.
The MS includes a system of precision tracking chambers up to \(|\eta| = 2.7\) and fast detectors for triggering up to \(|\eta| = 2.4\).
The luminosity is measured mainly by the LUCID--2~\cite{LUCID2} detector, which is located close to the beampipe.
A two-level trigger system is used to select events~\cite{TRIG-2016-01}.
The first-level trigger is implemented in hardware and uses a subset of the detector information
to accept events at a rate below \qty{100}{\kHz}.
This is followed by a software-based trigger that
reduces the accepted event rate to \qty{1}{\kHz} on average
depending on the data-taking conditions.
A software suite~\cite{SOFT-2022-02} is used in data simulation, in the reconstruction
and analysis of real and simulated data, in detector operations, and in the trigger and data acquisition
systems of the experiment.


\section{Data and simulated event samples} \label{sec:DataAndMCs}
\subsection{Simulated samples for the development of the method} \label{sec:DevelopmentMC}
Monte Carlo (MC) simulated event samples~\cite{SOFT-2010-01}
of signal and backgrounds are used to develop
the $\tauhadmurm$ method and to evaluate the $\tauhad$ reconstruction and identification efficiencies,
as well as the background rejection power. The signal samples consist of BSM gravitons~\cite{Graviton_theory}
decaying into a pair of Higgs bosons, with the hypothetical graviton mass ranging from
1~$\TeV$ to 5~$\TeV$.
For this study, the SM Higgs bosons are constrained to decay into
a pair of $\tau$-leptons. The signal process is referred to as $\GHHFourtau$.

High-\pT, semileptonically decaying heavy-flavour hadrons may produce a detector signature that has some similarities with
the signal sample of boosted $\tmth$ pairs, since the invariant mass of the charm hadron produced in the
semileptonic decay of a bottom hadron is comparable to the $\tau$-lepton mass.
The top-antitop-quark process ($\ttbar$) is used to model this type of background.

The $\GHHFourtau$ signal samples were simulated using the
\MGNLO~\cite{Alwall:2014hca} generator using matrix elements (ME)
at leading-order (LO) in quantum chromodynamics (QCD)
with the \NNPDF[2.3lo]~\cite{Ball:2012cx}
parton distribution function (PDF) set.
The production of \ttbar events was modelled using the
\POWHEGBOX[v2]~\cite{Frixione:2007nw,Nason:2004rx,Frixione:2007vw,Alioli:2010xd}
generator at next-to-leading order (NLO) in QCD with the \NNPDF[3.0nlo]~\cite{Ball:2014uwa} PDF set
and the \hdamp parameter\footnote{The
\hdamp parameter is a resummation damping factor and one of the
parameters that controls the matching of \POWHEG matrix elements to
the parton shower and thus effectively regulates the
high-transverse-momentum radiation against which the \ttbar system recoils.} set
to 1.5~\mtop ~\cite{ATL-PHYS-PUB-2016-020}.

\PYTHIA[8.230]~\cite{Sjostrand:2014zea} with the A14~\cite{ATL-PHYS-PUB-2014-021} set of tuned parameters (tune) and \NNPDF[2.3lo]
PDF set was used for the simulation of the parton showering and hadronisation for both the $\GHHFourtau$ samples and
the \ttbar samples. The decays of bottom and charm hadrons were performed by \EVTGEN[1.6.0]~\cite{Lange:2001uf}.
The effect of multiple interactions in the same and neighbouring bunch crossings (pile-up) was modelled
by overlaying the original hard-scattering
event with simulated inelastic proton--proton collisions generated by
\PYTHIA[8.186]~\cite{Sjostrand:2007gs} with the A3 tune~\cite{ATL-PHYS-PUB-2016-017}
and the \textsc{MSTW2008lo} PDF set~\cite{Martin:2009iq}.
The MC samples were reweighted so that the pile-up distribution matches the one observed in the data.
All MC samples were passed  through the
ATLAS detector simulation based on \GEANT~\cite{Agostinelli:2002hh}.

\subsection{Data and simulated samples for the validation of the method} \label{sec:Benchmark}
For the $\Zttmuhad$ validation analysis,
the Run~2 data sample collected in proton--proton
collisions at the LHC with $\sqrt{s}=13~\TeV$
and a 25~ns bunch crossing interval is used.
The integrated luminosity of the sample recorded
while all relevant components of the ATLAS detector were operating
normally  corresponds to 140~\ifb~\cite{DAPR-2018-01,DAPR-2021-01}.

Simulated samples provide predictions for both signal and background processes.
As described in Section~\ref{sec:DevelopmentMC},
the simulation includes the effect of pile-up and the detector response.
All MC samples undergo calibrations and corrections to match the performance in data.
\EVTGEN[1.6.0]~\cite{Lange:2001uf} was used again to decay bottom and charm hadrons.
The rest of this section gives details of the specific MC samples used in this study.

The dominant production channel of the boosted $\Zttmuhad$ final states,
a $Z$ boson produced in association with jets, was modelled using the
\SHERPA[2.2.14]~\cite{Bothmann:2019yzt} generator for the $\Ztautau$ channel,
while \SHERPA[2.2.11] was used to model $Z\rightarrow \ell\ell~(\ell=e,\mu)$ decays.
The ME calculations ranged from NLO in QCD for final
states with up to two additional parton emissions to LO in QCD
for up to five additional parton emissions.
The MEs were merged with the \SHERPA parton shower following the \MEPSatLO~\cite{Hoeche:2009rj} prescription and using the \NNPDF[3.0nnlo]~\cite{Ball:2014uwa} PDF set.
The production of $W$ bosons with jets, and the production of $WW, WZ$, and $ZZ$ boson pairs, were simulated using \SHERPA[2.2.11] with configurations similar to those used in the $Z$+jets sample.
Measurements of the distribution of $Z$-boson transverse momentum~($\pt^{Z}$)
using light-lepton pair events at $\sqrt{s}=13~\TeV$~\cite{STDM-2018-14} indicate that
\SHERPA[2.2.11] underestimates the cross-section for $\pt^{Z} > 100~\GeV$
by approximately 10\%~\cite{PMGR-2021-01}.
The simulation of initial-state QCD
radiation is expected to be the same in the \SHERPA[2.2.14] samples used in this study. A 10\%
correction is therefore applied to the
predicted event yields for $\Ztautau$ in the validation analysis,
with the full size of this correction quoted as a systematic uncertainty.

The production of \ttbar events was modelled using
the same configuration as the one used for the method development as in Section~\ref{sec:DevelopmentMC}.
Single-top \(t\)-channel production was modelled using the
\POWHEGBOX[v2]~\cite{Frederix:2012dh,Nason:2004rx,Frixione:2007vw,Alioli:2010xd}
generator at NLO in QCD using the four-flavour scheme and the
corresponding \NNPDF[3.0nlo] PDF set.
The associated production of top quarks with \(W\) bosons (\(tW\)) was
modelled using the
\POWHEGBOX[v2]
generator at NLO in QCD using the five-flavour scheme and the
\NNPDF[3.0nlo] PDF set.
The diagram removal scheme~\cite{Frixione:2008yi} was used to
remove interference and overlap with \ttbar production.
In both cases the generated events were
interfaced to \PYTHIA[8.230] using the A14
tune and the \NNPDF[2.3lo] PDF set.

The production of Higgs bosons via gluon--gluon fusion and
vector--boson fusion was modeled following the configuration detailed
in Ref.~\cite{HIGG-2019-09}.

\section{Development of the boosted $\tauhadmurm$ reconstruction method} \label{sec:murm}
\subsection{Muon removal and re-reconstruction of the $\tauhad$ candidate} \label{sec:muon_removal}
The $\tauhadmurm$ method was developed
with simulated $\GHHFourtau$ signal samples
with generator-level information used to evaluate
the performance of the method.
The angular distance at the generator-level between the $\tauhadvis$ and the muon is measured by $\dRtttruthvis$.
In Figure~\ref{fig:dr}, the $\dRtttruthvis$ distributions of the
\tmth pairs in the \GHHFourtau samples are presented.
\begin{DIFnomarkup}\begin{figure}[btp]
\begin{center}
\includegraphics[width=0.40\textwidth]{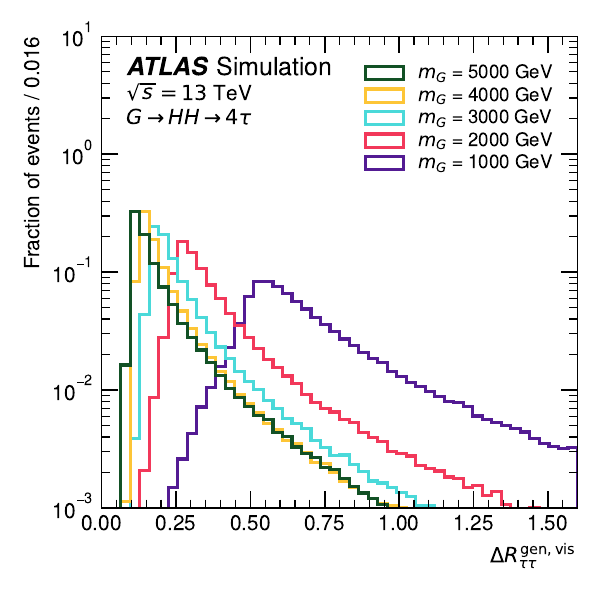}
\caption{The distributions of the $\dRtttruthvis$ for the generator-level
\tmth pair in \GHHFourtau signal events, with
graviton masses $m_G$ ranging from 1000~\GeV to 5000~\GeV.
}
\label{fig:dr}
\end{center}
\end{figure}\end{DIFnomarkup}
For $\dRtttruthvis < 0.4$, it is more likely that the muon will
be reconstructed inside the $\tauseed$ jet.

In this study, MC generator-level information is used to match generated $\tmth$ pairs with
their reconstructed counterparts. The reconstructed $\tauseed$ jet is generator-matched if at least one
ID track matches the muon from the $\taulep$ decay or the charged hadron(s) from the $\tauhad$ decay.
Only $\tauseed$ jets originating from a boosted $\tmth$ pair are considered.
Other $\tauseed$ jets, such as those associated with QCD jets, are neglected.
The MC generator-level visible $\tau$-lepton decay products are required to satisfy the
requirements $\pt > 20~\GeV$ and $|\eta| < 2.5$ to ensure that the $\tauseed$ jets
are within the tracking acceptance of the ATLAS detector.

\begin{DIFnomarkup}\begin{figure}[btp]
\begin{center}
\subfloat[]{\includegraphics[width=0.4\textwidth]{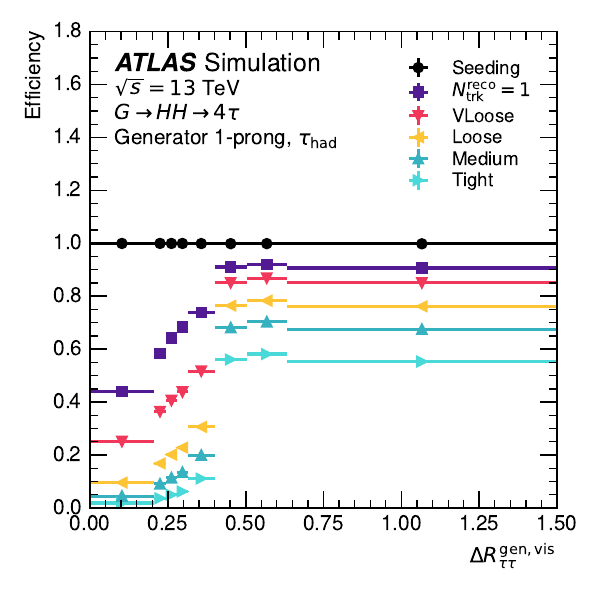}\label{fig:effi_off_1p}}
\subfloat[]{\includegraphics[width=0.4\textwidth]{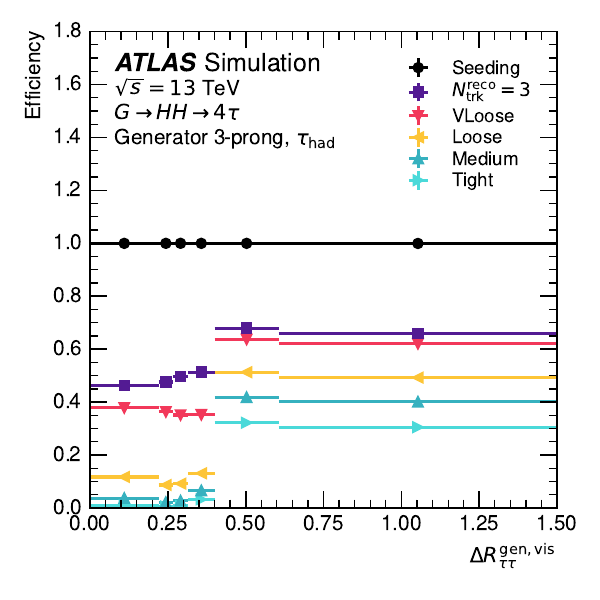}\label{fig:effi_off_3p}}
\caption{The combined reconstruction and identification efficiencies of the standard ATLAS TauID for
generator-level (a) 1-prong and (b) 3-prong $\tmth$ pairs
at all working points as a function of $\dRtttruthvis$. The `Seeding' line
shows the efficiency of a $\tauseed$ jet that matches a generator-level $\tmth$ pair being
reconstructed; the `$N_\mathrm{trk}^\mathrm{reco}=1\text{ or }3$' line shows the efficiency of a generator-level $\tauhad$ being
reconstructed with the correct number of associated charged-particle tracks.
The `VLoose', `Loose', `Medium', and `Tight' lines show the efficiencies of the
TauID working points.
}
\label{fig:effi_off}
\end{center}
\end{figure}\end{DIFnomarkup}
Below the $\dRtttruthvis < 0.4$ threshold, the standard $\tauhad$ reconstruction and identification efficiencies drop significantly.
Figure~\ref{fig:effi_off} shows the combined reconstruction and identification efficiencies of
the standard ATLAS $\tauhad$ reconstruction and TauID algorithms as a function of $\dRtttruthvis$ for generator-level 1-prong and 3-prong $\tauhad$ in
$\tmth$ pairs. Here `n-prong' denotes the number of charged hadrons that originate from the $\tauhad$ decay.
The `$N_\mathrm{trk}^\mathrm{reco} = 1\text{ or }3$' lines shows the efficiency of a generator-level 1-prong or 3-prong $\tauhad$ being
reconstructed with the correct number of associated charged-particle tracks.
The efficiency for a $\tauseed$ jet to be reconstructed from a $\tmth$ pair remains high for all
values of $\dRtttruthvis$. However, the efficiencies for all standard TauID working
points drop significantly in both the 1-prong and 3-prong cases when $\dRtttruthvis < 0.4$.
The tighter the working point, the greater the fractional loss in efficiency due to the presence of the nearby
muon. The performance of the standard TauID algorithm is set as the baseline in this study.

The muon's nature as a minimum ionising particle, combined with the fact
that its reconstruction is independent of its isolation~\cite{MUON-2018-03},
provides a motivation for excluding the ID track and calorimeter
clusters associated with any nearby muon from the standard ATLAS $\tauhad$ reconstruction algorithm.
By removing these contributions, the $\tauseed$ jet would better represent
only the $\tauhadvis$, improving the accuracy of the TauID algorithm.
Specifically, the ID track and clusters associated with a reconstructed muon that
satisfies the `Medium' muon identification working point
are removed if the track or cluster is found within the reconstructed $\tauseed$ jet.
For calorimeter energy clusters, the algorithm checks if the energy of the cluster
associated with the muon is compatible with the
expected energy loss~\cite{MUON-2022-01} from its interactions with the calorimeters between the ID and the MS
and removes the cluster only when this is the case.
After the muon removal, the standard
$\tauhad$ reconstruction algorithm is re-run on the $\tauseed$ jet. In this way, the relevant TauID variables for
the $\tauhad$ candidate can be calculated without the muon component.

\subsection{Performance of the method} \label{sec:murm_perf}
\begin{DIFnomarkup}\begin{figure}[hbtp]
\begin{center}
\subfloat[]{\includegraphics[width=0.4\textwidth]{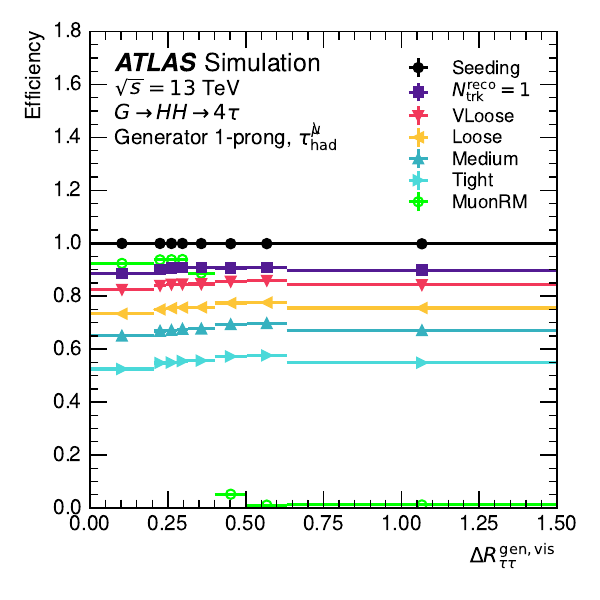}\label{fig:effi_new_1p}}
\subfloat[]{\includegraphics[width=0.4\textwidth]{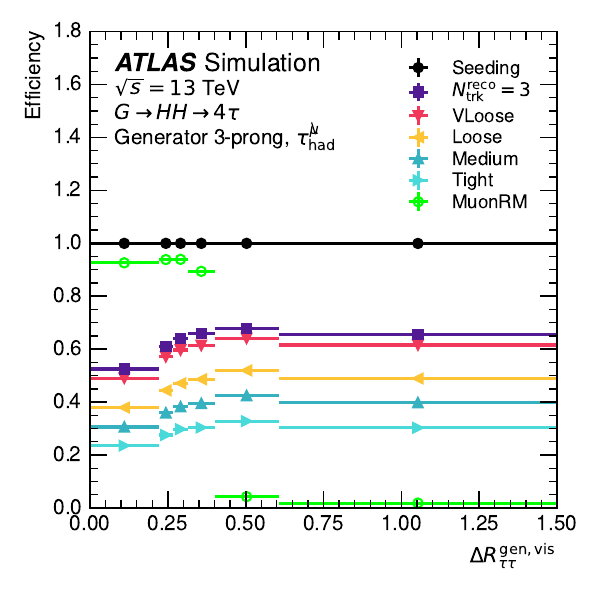}\label{fig:effi_new_3p}}
\caption{The combined reconstruction and TauID efficiencies
after the muon removal for generator-level (a) 1-prong
and (b) 3-prong
$\tmth$ pairs at all working points as a function of $\dRtttruthvis$.
The `Seeding' line
shows the efficiency of a $\tauseed$ jet that matches a generator-level $\tmth$ pair being
reconstructed; the `$N_\mathrm{trk}^\mathrm{reco}=1\text{ or }3$' line shows the efficiency of a generator-level $\tauhad$ being
reconstructed with the correct number of associated charged-particle tracks.
The `VLoose', `Loose', `Medium', and `Tight' lines show the efficiencies of the
TauID working points.
The `MuonRM' line shows the efficiency of a muon being identified and removed from the $\tauseed$ jet.
The transitions of the `MuonRM' line
between $0.35 < \dRtttruthvis < 0.45$ is due to limited detector
resolution in the direction of the reconstructed $\tauhadvis$ and muon.
}
\label{fig:effi_new}
\end{center}
\end{figure}\end{DIFnomarkup}
Figure~\ref{fig:effi_new} shows the combined reconstruction and identification efficiencies after the muon removal as a function of
$\dRtttruthvis$ for generator-level 1-prong and 3-prong $\tmth$ pairs for all working points.
The reconstruction and identification efficiencies after the muon removal show a considerable improvement compared to those shown
in Figure~\ref{fig:effi_off}. The signal efficiency is recovered almost completely for every working point
for both 1-prong and 3-prong $\tauhad$. In the 3-prong cases, the efficiency of TauID is limited by the
accurate reconstruction of the correct number of tracks associated with the highly boosted
$\tauhad$ candidates.

Roughly 95\% of the $\tauseed$ jets have a muon removed in the region where $\dRtttruthvis < 0.4$.
This can be expected given the 97\% efficiency of the `Medium' muon working point,
with a dip below 90\% in the region where the generator-level $|\eta|$ of the $\tauhadvis$ is $\absetatruthvis < 0.1$. This is illustrated in
Figure~\ref{fig:etaRM}, which shows the signal efficiencies of the TauID working points after the muon removal as a
function of $\absetatruthvis$ for generator-level 1-prong and 3-prong $\tauseed$ jets with
$\dRtttruthvis < 0.4$.
The slight decrease in muon removal efficiency in the low $\absetatruthvis$ region is
caused by the non-instrumented regions of the MS.

\begin{DIFnomarkup}\begin{figure}[hbtp]
\begin{center}
\subfloat[]{\includegraphics[width=0.40\textwidth]{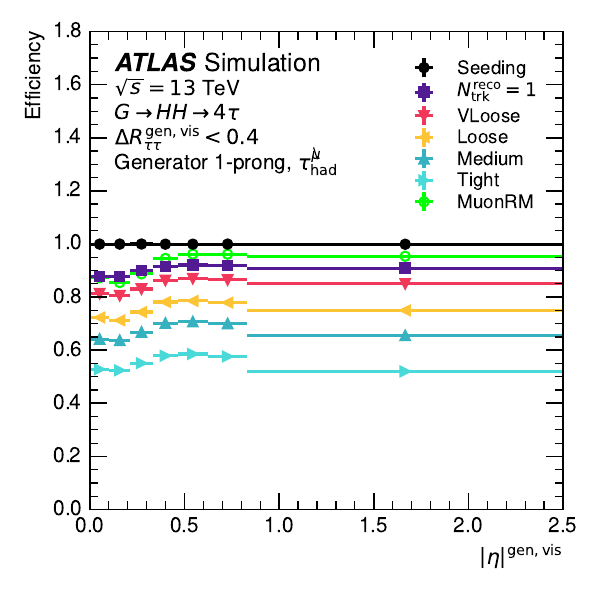}\label{fig:etaRM_1p}}
\subfloat[]{\includegraphics[width=0.40\textwidth]{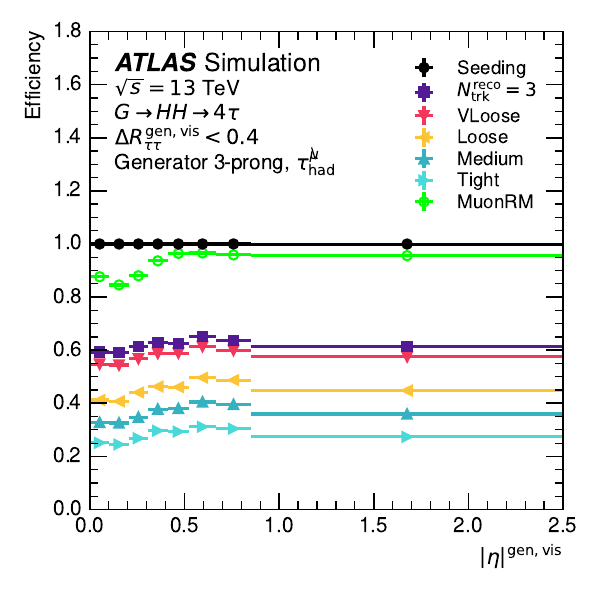}\label{fig:etaRM_3p}}
\caption{The combined reconstruction and TauID efficiencies
after the muon removal for generator-level (a) 1-prong
and (b) 3-prong
$\tmth$ pairs at all working points as a function of $\absetatruthvis$ for $\tauseed$ jets
with $\dRtttruthvis < 0.4$.
The `Seeding' line shows the efficiency of a $\tauseed$ jet that matches a generator-level $\tmth$ pair being reconstructed;
the `$N_\mathrm{trk}^\mathrm{reco}=1\text{ or }3$' line shows the efficiency of a generator-level $\tauhad$ being reconstructed with the correct number of associated charged-particle tracks.
The `VLoose', `Loose', `Medium', and `Tight' lines show the efficiencies of the TauID working points.
The `MuonRM' line shows the efficiency of a muon being identified and removed from the $\tauseed$ jet.
}
\label{fig:etaRM}
\end{center}
\end{figure}\end{DIFnomarkup}

The signal identification efficiencies at all working points of the standard ATLAS TauID are tuned to show minimum
dependency on the generator-level $\pt$ of the $\tauhadvis$ ($\pttautruthvis$) and pile-up~\cite{ATL-PHYS-PUB-2022-044}. The stability
of the TauID working point efficiencies, after the muon removal, against these variables are shown in Figure~\ref{fig:stability}.
To focus on the objects of interest, only generator-level $\tmth$ pairs with the muon identified inside the $\tauseed$ jet and removed ($\dRttreco < 0.4$)
are included in these plots. For comparison, the TauID working point efficiencies as a function of the same variables
for $\tauseed$ objects in which muon removal is not required ($\dRtttruthvis > 0.45$)
are shown in Figure~\ref{fig:45drstability}. Similar behaviour is observed in Figures~\ref{fig:stability}~and~\ref{fig:45drstability}.
The improved performance and good stability across different working points demonstrate that, after the muon removal,
the TauID RNN receives as input the signal $\tauseed$ jet as if it were a $\tauhad$ that is isolated (free from interference of surrounding particles).

\begin{DIFnomarkup}\begin{figure}[hbtp]
\begin{center}
\subfloat[]{\includegraphics[width=0.40\textwidth]{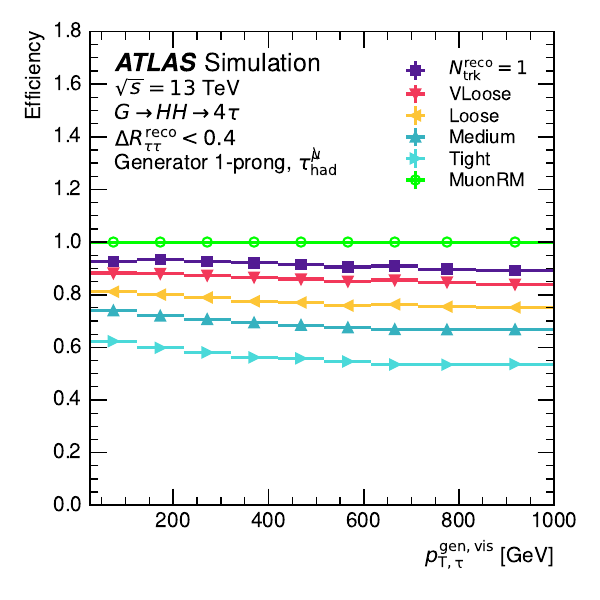} \label{fig:ptDR04_1p}}
\subfloat[]{\includegraphics[width=0.40\textwidth]{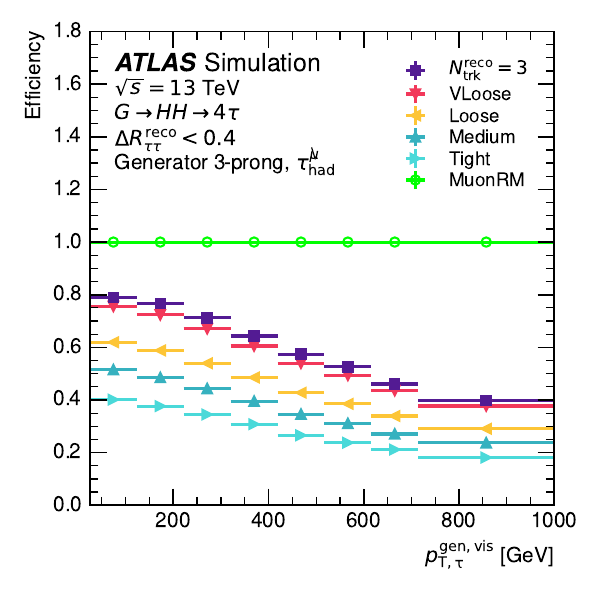} \label{fig:ptDR04_3p}}
\\
\subfloat[]{\includegraphics[width=0.40\textwidth]{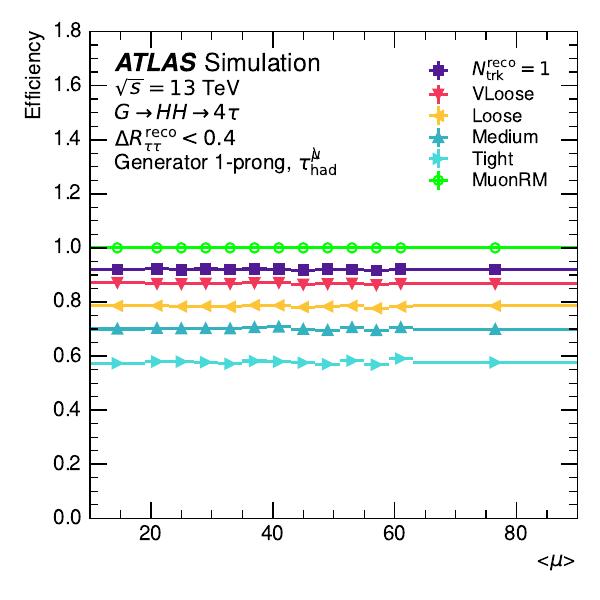} \label{fig:muDR04_1p}}
\subfloat[]{\includegraphics[width=0.40\textwidth]{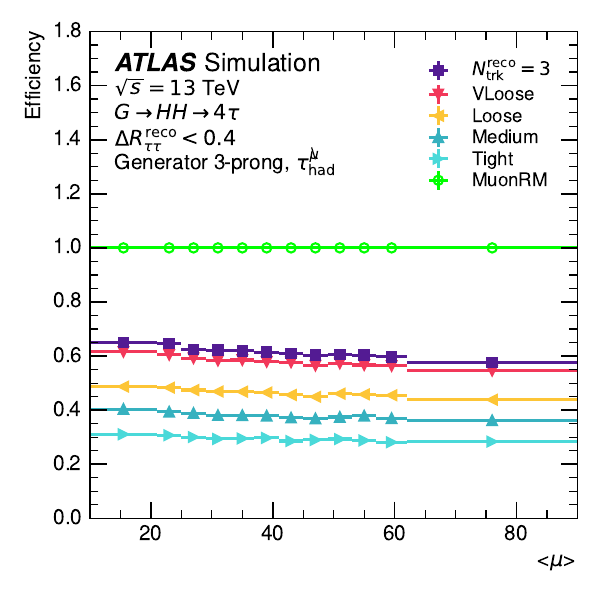} \label{fig:muDR04_3p}}
\caption{
The signal efficiencies of the TauID working points as a function of
(a, b) the generator-level transverse momentum $\pttautruthvis$
and (c, d)
the average number of interactions per bunch crossing <$\mu$>
for generator-level
1-prong and 3-prong $\tauseed$ jets when the muon is identified inside the $\tauseed$ jet and removed ($\dRttreco<0.4$).
The `$N_\mathrm{trk}^\mathrm{reco}=1\text{ or }3$' line shows the efficiency of a generator-level $\tauhad$ being
reconstructed with the correct number of associated charged-particle tracks.
The `VLoose', `Loose', `Medium', and `Tight' lines show the efficiencies of the
TauID working points.
The `MuonRM' line shows the efficiency of a muon being identified and removed from the $\tauseed$, and is by definition 1.0 for all subfigures.
}
\label{fig:stability}
\end{center}
\end{figure}\end{DIFnomarkup}

\begin{DIFnomarkup}\begin{figure}[hbtp]
\begin{center}
\subfloat[]{\includegraphics[width=0.40\textwidth]{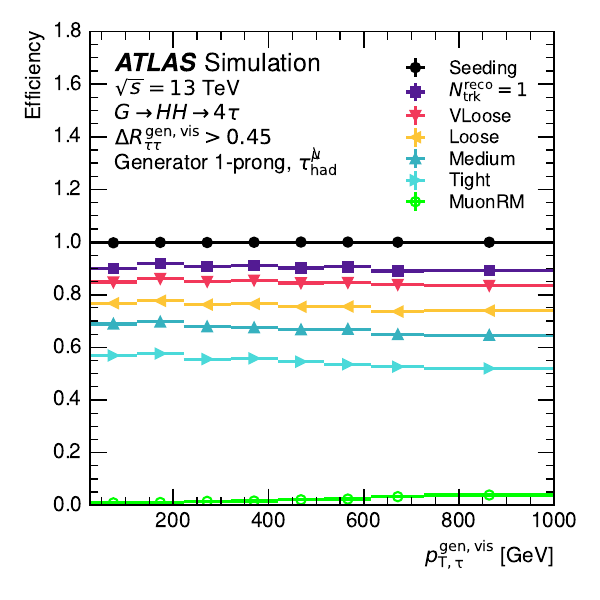}  \label{fig:pt45dr_1p}}
\subfloat[]{\includegraphics[width=0.40\textwidth]{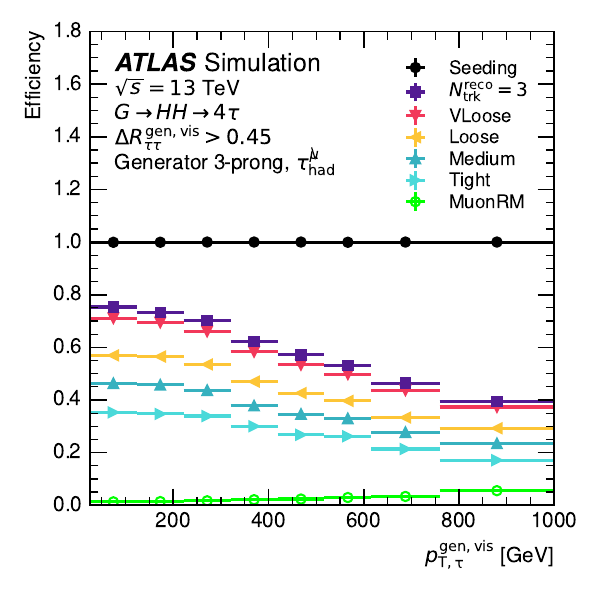}  \label{fig:pt45dr_3p}}
\\
\subfloat[]{\includegraphics[width=0.40\textwidth]{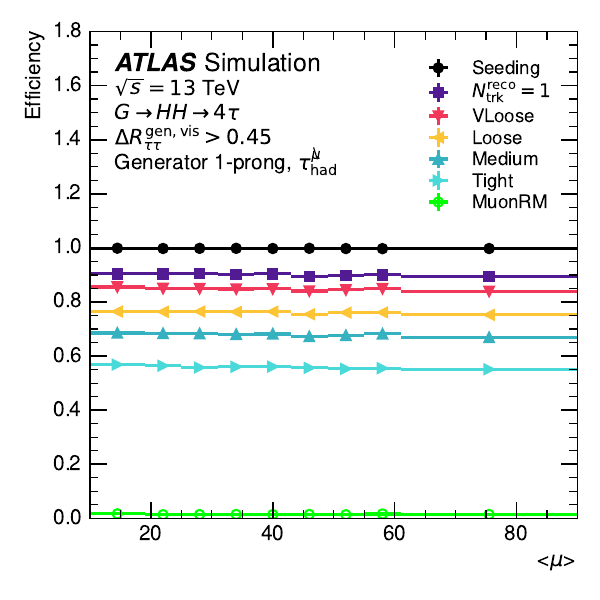}  \label{fig:mu45dr_1p}}
\subfloat[]{\includegraphics[width=0.40\textwidth]{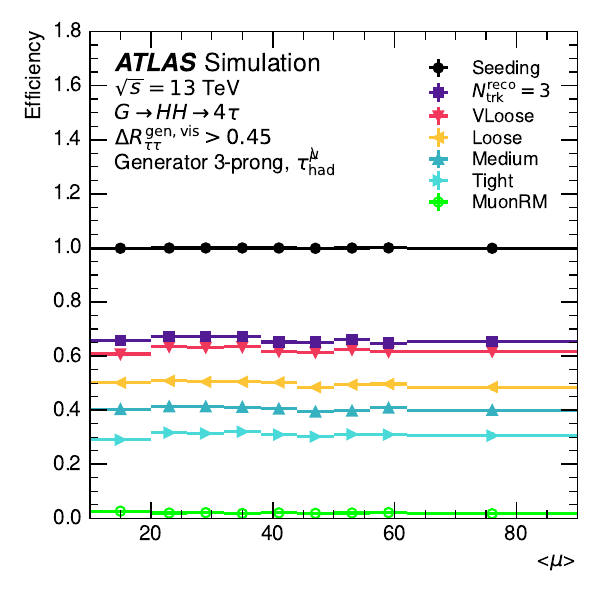}  \label{fig:mu45dr_3p}}
\caption{
The signal efficiencies of the TauID working points as a function of
(a, b) the generator-level $\pttautruthvis$
and (c, d)
the average number of interactions per bunch crossing (<$\mu$>)
for the generator-level 1-prong and 3-prong $\tauseed$ jets with $\dRtttruthvis > 0.45$.
In this region, the muon removal should not affect the results.
The `Seeding' line
shows the efficiency of a $\tauseed$ jet that matches to a generator-level $\tmth$ pair being
reconstructed; the `$N_\mathrm{trk}^\mathrm{reco}=1\text{ or }3$' line shows the efficiency of a generator-level $\tauhad$ being
reconstructed with the correct number of associated charged-particle tracks.
The `VLoose', `Loose', `Medium', and `Tight' lines show the efficiencies of the
TauID working points.
The `MuonRM' line shows the efficiency of a muon being identified and removed from the $\tauseed$ jet,
and is expected to be low in high $\dRtttruthvis$ region.
}
\label{fig:45drstability}
\end{center}
\end{figure}\end{DIFnomarkup}

After the removal of the overlapping muon, the precision with which the four-momentum of the $\tauseed$ jet is reconstructed
improves significantly. Figure~\ref{fig:reso_eta_allp} shows the distributions of
the difference between the generated and reconstructed $\eta$ (residuals) before and after the muon removal.
Compared to the performance before muon removal,
the $\eta$ residuals that correspond to the 68\% percentile (core resolution)
improves by a factor of 15.
The distributions of the $\tauhadvis$ $\pt$ residuals ($\pt / \pt^\mathrm{gen}$) are shown in Figure~\ref{fig:reso_pt_allp}.
These results agree well with those reported for isolated $\tauhad$ in Ref.~\cite{PERF-2014-06},
further demonstrating the effectiveness of the muon removal method.

\begin{DIFnomarkup}\begin{figure}[hbtp]
\begin{center}
\subfloat[]{\includegraphics[width=0.4\textwidth]{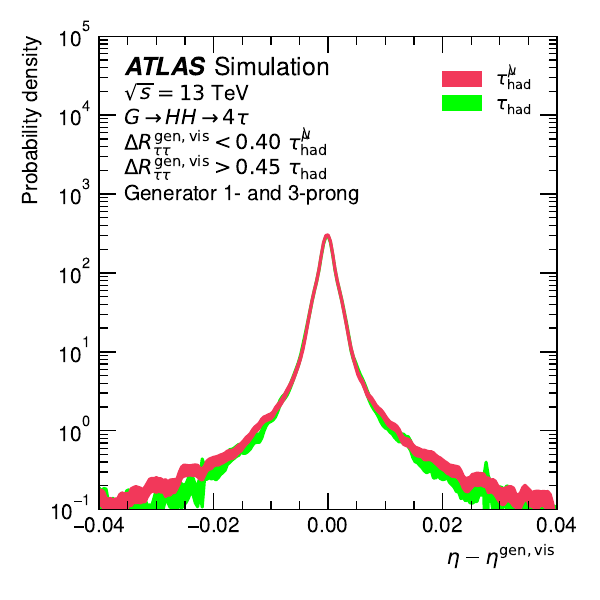} \label{fig:reso_eta_allp}}
\subfloat[]{\includegraphics[width=0.4\textwidth]{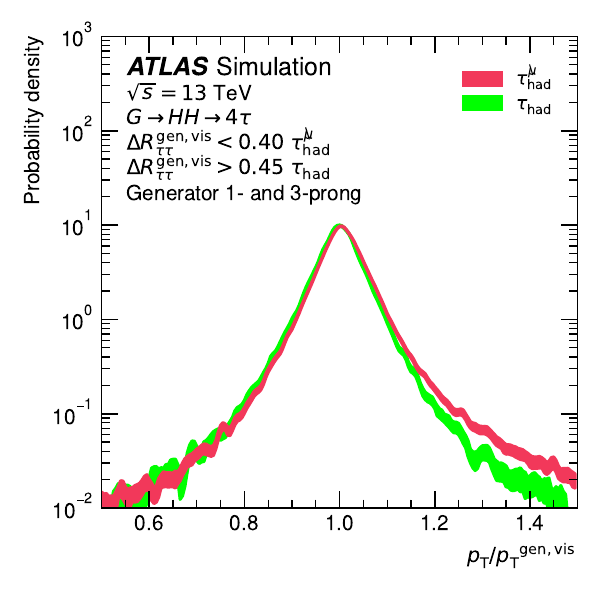}  \label{fig:reso_pt_allp}}
\caption{The distributions of the residuals for
(a) $\eta$ and
(b) $\pt$ after calibrations
with standard $\tauhad$ candidates and $\dRtttruthvis > 0.45$
and with $\tauhadmurm$ candidates and $\dRtttruthvis < 0.4$.
The bands represent the statistical uncertainties.}
\label{fig:resolutions}
\end{center}
\end{figure}\end{DIFnomarkup}

Having demonstrated that the reconstruction and identification
efficiencies are significantly improved by the $\tauhadmurm$ method,
the background rejection power is studied.
The production of \ttbar events is considered as a source of
high-$\pt$ heavy-flavour jets, which represents an example background to the $\tauhadmurm$ signal.
The background rejection at the `Medium' TauID working point,
which is defined as the ratio of the total number of reconstructed $\tauhad$ candidates
before muon removal to the number of false positives before or after muon removal,
are shown in Figure~\ref{fig:rej},
as functions of the reconstructed $\tauhadvis~\pt$. For the background rejection figures,
the event selections are mostly based on the reconstructed properties instead of the generator-level
information. A $\tauseed$ jet reconstructed in the background sample is required to have reconstructed
$20~\GeV<\pt<300~\GeV$,
$|\eta|<2.5$,
and to not be generator-matched to a $\tauhad$ from the semileptonic decay of a bottom or top quark.
In addition, a reconstructed muon is required to be found inside the $\tauseed$ jet.
The $\tauhad$ candidates in which no muon is present are not considered in Figure~\ref{fig:rej}, as identical
TauID results would be expected for these cases.

\begin{DIFnomarkup}\begin{figure}[hbtp]
\begin{center}
\subfloat[]{\includegraphics[width=0.4\textwidth]{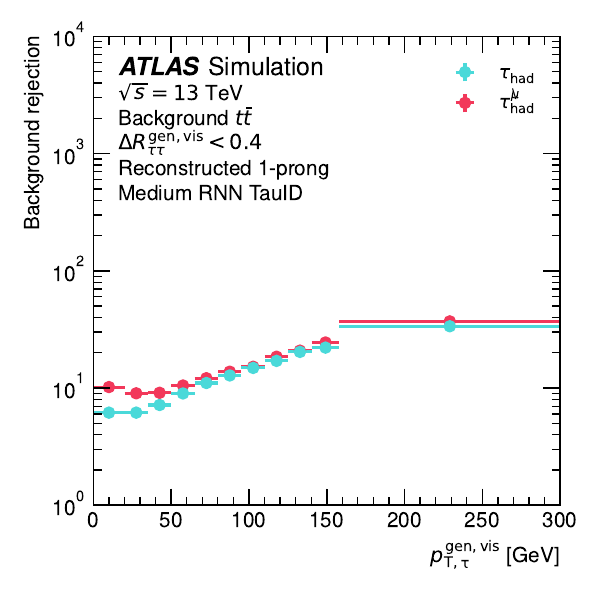}  \label{fig:rej_pt_1p}}
\subfloat[]{\includegraphics[width=0.4\textwidth]{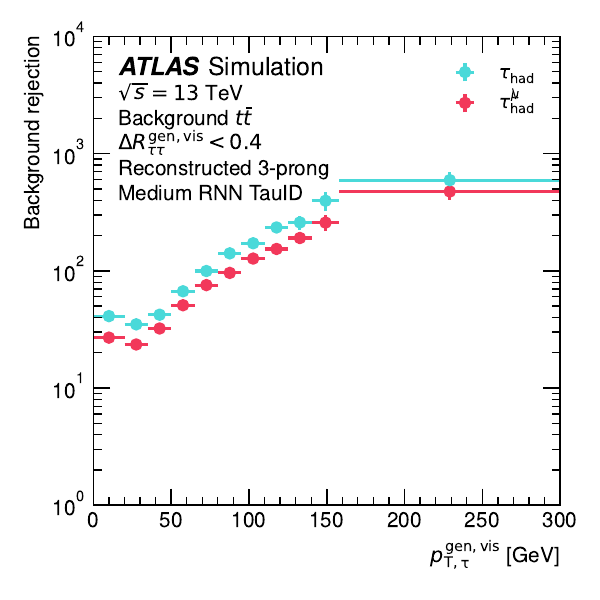}  \label{fig:rej_pt_3p}}
\caption{
The background rejection
for $\tauseed$ jets originating from
semileptonic $b$-hadron decays in \ttbar events at the `Medium' TauID working point,
as a function of the reconstructed \pt for
(a) the reconstructed 1-prong candidates and
(b) the reconstructed 3-prong candidates.
}
\label{fig:rej}
\end{center}
\end{figure}\end{DIFnomarkup}
\begin{DIFnomarkup}\begin{figure}[hbtp]
\begin{center}
\subfloat[]{\includegraphics[width=0.4\textwidth]{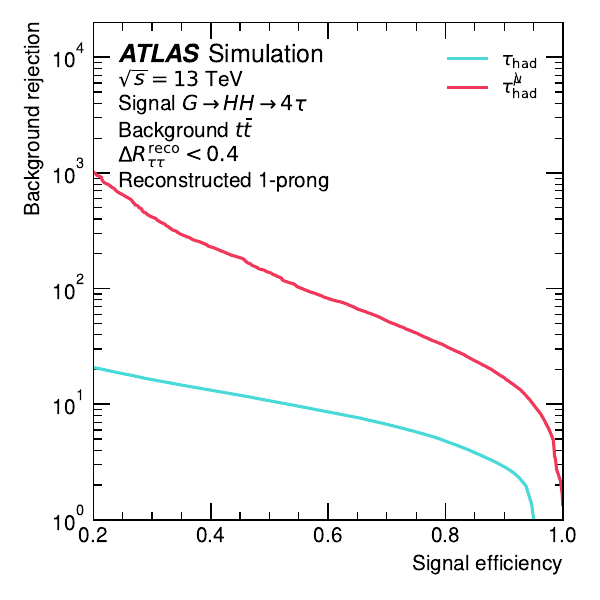}\label{fig:roc_1p}}
\subfloat[]{\includegraphics[width=0.4\textwidth]{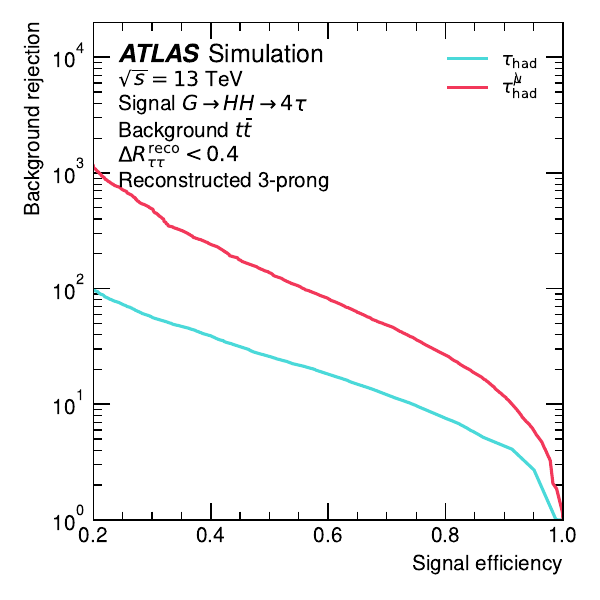}\label{fig:roc_3p}}
\caption{
Background rejection as a function of signal efficiency
with and without muon removal for
(a) the reconstructed 1-prong candidates and
(b) the reconstructed 3-prong candidates.
For the background sample, $\ttbar$ events are used.
The combination of all $\GHHFourtau$ samples with
$m_G$ ranging from 1~\TeV to
5~\TeV is used as the signal sample.
}
\label{fig:roc}
\end{center}
\end{figure}\end{DIFnomarkup}

It is likely that after removing the muon, the
number of reconstructed charged-particle tracks inside
the $\tauseed$ jet decreases by one compared
to the standard reconstruction.
Thus the background compositions of $\tauhad$ candidates reconstructed with the $\tauhadmurm$ method
are different from the standard $\tauhad$ candidates.
The background rejection power decreases slightly in the 3-prong case compared to the standard TauID algorithm.
Without the presence of the muon within the $\tau_\mathrm{seed}$ in the case of background events,
the TauID algorithm finds it more challenging to reject semileptonic heavy-flavour jets.
The 1-prong background rejection power increases slightly in
the low-\pt region after removal of the muon due to fewer $\tauhad$ candidates being reconstructed after the muon removal.

The signal efficiency observed in the \GHHFourtau samples and the background rejection power observed in the $\ttbar$ sample
are combined to form the so-called receiver operating characteristic (ROC) curves.
In addition to the same reconstruction level selections as the background $\tauseed$ jets discussed previously, the
$\tauseed$ jets in the signal samples are required to be generator-matched to the $\tmth$ pairs. To minimise the
bias introduced by the misalignment of the signal and background momentum spectra, the background events
are reweighted so that the \pt distribution of the background sample matches that of the signal samples.
The ROC curves illustrating the performance with and without muon removal are shown in Figure~\ref{fig:roc}.
An order-of-magnitude performance gain is seen across the spectrum for both the 1-prong and 3-prong cases
when the muon removal is applied.

To demonstrate the performance of the $\tauhadmurm$ method in reconstructing and
identifying individual di-$\tau$ systems within the \GHHFourtau process,
Figure~\ref{fig:effi_mass} presents the combined $\tauhad$ reconstruction and
identification efficiencies as a function of the generated graviton mass.
The efficiencies correspond to the identification of individual di-$\tau$ systems originating from Higgs boson decays.

Compared to the standard ATLAS TauID algorithm, the $\tauhadmurm$ method demonstrates a complete recovery
of the identification efficiency for both the 1-prong and 3-prong $\tauhad$ decays for high-mass $\GHHFourtau$ events,
to the level expected for isolated $\tauhad$ candidates reconstructed with the standard TauID algorithm.
\begin{DIFnomarkup}\begin{figure}[hbtp]
\begin{center}
\subfloat[]{\includegraphics[width=0.40\textwidth]{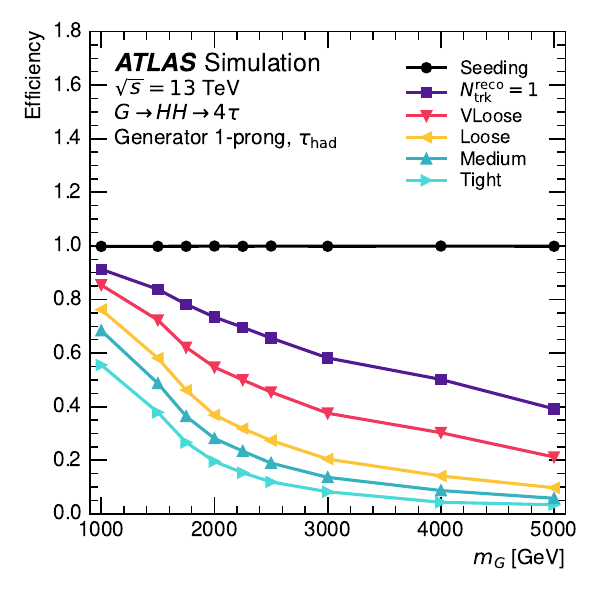} \label{fig:effi_mass_std_1p}}
\subfloat[]{\includegraphics[width=0.40\textwidth]{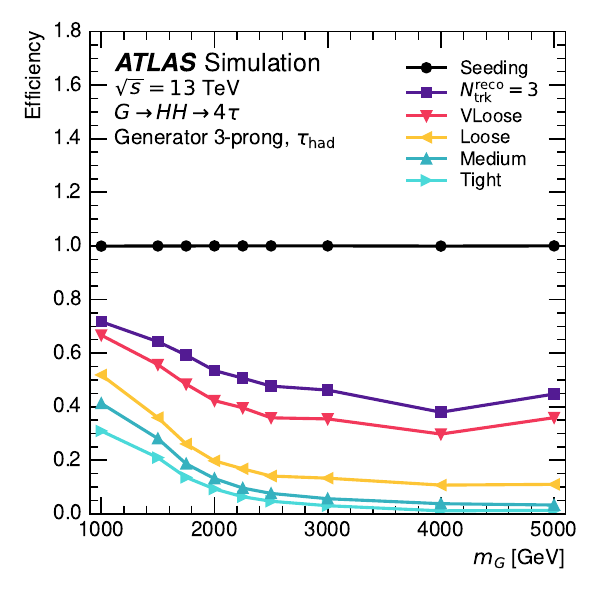} \label{fig:effi_mass_std_3p}}
\\
\subfloat[]{\includegraphics[width=0.40\textwidth]{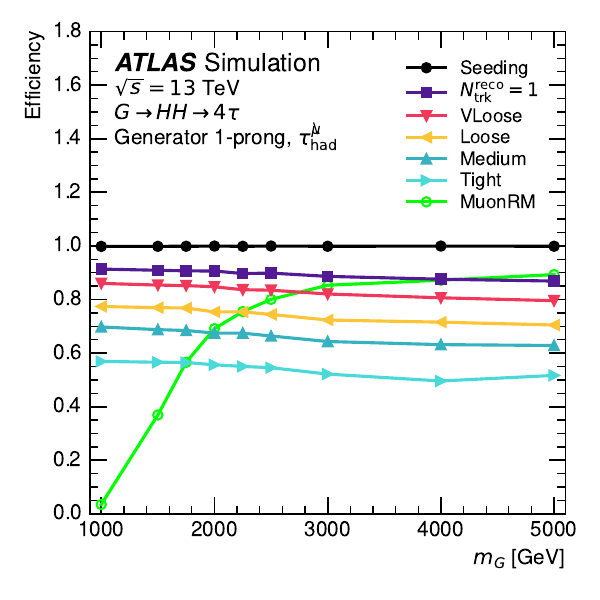}\label{fig:effi_mass_new_1p}}
\subfloat[]{\includegraphics[width=0.40\textwidth]{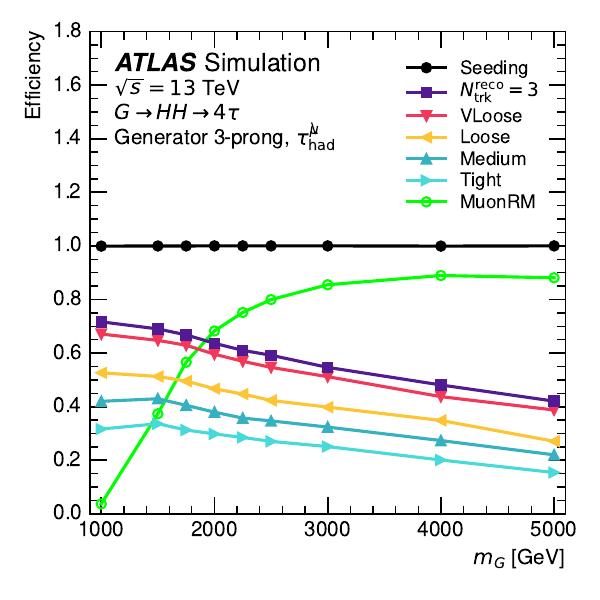}\label{fig:effi_mass_new_3p}}
\caption{The signal efficiencies of the TauID working points, as a function of $m_G$,
for generator-level (a) 1-prong and (b) 3-prong $\tauhad$ with the standard ATLAS TauID algorithm
and for generator-level (c) 1-prong and (d) 3-prong $\tauhad$ with the $\tauhadmurm$ method.
The efficiencies shown correspond to the identification of individual di-$\tau$ systems originating from Higgs boson decays.
The `Seeding' line
shows the efficiency of a $\tauseed$ jet that matches a generator-level $\tmth$ pair being
reconstructed. The `$N_\mathrm{trk}^\mathrm{reco}=1\text{ or }3$' line shows the efficiency of a generator-level $\tauhad$ being
reconstructed with the correct number of associated charged-particle tracks.
The `VLoose', `Loose', `Medium', and `Tight' lines show the efficiencies of the
TauID working points.
The `MuonRM' line shows the efficiency of a muon being identified and removed from the $\tauseed$ jet.
}
\label{fig:effi_mass}
\end{center}
\end{figure}\end{DIFnomarkup}

\FloatBarrier

\section{Validation of the $\tauhadmurm$ method in the $\Zttmuhad$ final states} \label{sec:Zttbench}
The $\tauhadmurm$ method was developed for searches for high-mass BSM physics, such as the \GHH process,
as described in Section~\ref{sec:murm}. However, it is useful to validate its performance by considering a SM process
and data from proton-proton collisions.
In this paper, the Drell--Yan production of a $Z$ boson in association with high-$\pt$ jets from initial-state
QCD radiation is considered as a validation process.
\subsection{$Z$-boson kinematic reconstruction} \label{sec:collinear_assumptions_and_evt_reco}
The detector signature of the boosted $\Zttmuhad$ process includes one hadronically decaying $\tau$-lepton and a muon,
in association with significant missing transverse momentum ($\vecptmiss$) from the neutrinos produced in the two
$\tau$-lepton decays.
The $\vecptmiss$, with magnitude $\MET$, is estimated
as the negative vector sum of the transverse momentum
of all identified hard physics objects~\cite{JETM-2020-03}.
Tracks not associated with any such object are included in the soft term.

In this kinematic regime, the
$\tauhad$ and the muon are likely to be reconstructed within the same $\tauseed$ jet. In well measured events, the vector sum
of the transverse momenta of the three neutrinos dominates the measured $\vecptmiss$, which in the azimuthal direction lies between the observed muon and the $\tauhad$ candidate
in most cases. Also, on average, the $\vecptmiss$ should be closer in azimuth to the muon,
because the $\taulep$ decay produces two neutrinos and the $\tauhad$ decay only one.

Without incorporating $\vecptmiss$, reconstructing the $Z$-boson invariant mass is not possible. However, it is possible to approximate
the momenta of the neutrinos using the collinear approximation~\cite{ELLIS1988221} as follows:
\begin{itemize}
\item the transverse momenta of the three neutrinos dominate the measured $\vecptmiss$, and other contributions are negligible;
\item each $\tau$-lepton is sufficiently boosted such that the neutrino (or pair of neutrinos) produced in its decay is collinear with its visible decay products.
\end{itemize}
These assumptions allow the reconstruction of the momenta of the neutrino (or pair of neutrinos) produced in the decay
of each $\tau$-lepton.\footnote{In events in which the $\vecptmiss$ lies outside the azimuthal angle
between the visible decay products of the $\tau$-leptons, the
$\vecptmiss$ is projected onto the direction of the nearest visible decay, and the neutrino momentum
associated with the other $\tau$-lepton is set to zero. Furthermore, events for which collinear
reconstruction is not possible, i.e., with $\vecptmiss$ deviating by more than $90^{\circ}$ from the muon
or the $\tauhadmurm$ in $\deltaphi$, are discarded.}

Together with the momenta of the visible decay products, this procedure allows the momenta of
the two $\tau$-leptons, and hence the transverse momentum ($\ptcol$), and mass ($\mcol$)
of the system produced by the decay of the $Z$ boson, to be reconstructed.
The jets recoiling against the $Z$ boson are reconstructed using the anti-$k_t$ algorithm with a radius
parameter $R=0.4$ that uses particle flow objects~\cite{PERF-2015-09} as input.
These jets are then calibrated using simulation and data a to a precision of order 1\% for central jets.
\subsection{Event selection} \label{sec:ZttSelection}
Candidate events are required to have been
triggered by an un-prescaled single-muon trigger~\cite{TRIG-2018-01}
or an un-prescaled $\MET$ trigger~\cite{TRIG-2019-01}.
The thresholds of the $\pt$ required to fire each trigger varied
for different data-taking periods. For the single muon trigger,
the $\pt$ requirement for triggers selecting isolated muons
ranged from 20 to 26~$\GeV$, while the $\pt$ threshold
for muon trigger without isolation requirements remained
constant at 50~$\GeV$.
The \MET triggers had a threshold of 70~$\GeV$ for the 2015 data-taking period and remained constant at 110~$\GeV$ for the rest of Run~2.
Approximately 97\% of  $\Zttmuhad$ simulated events that satisfy the final signal selection criteria satisfy the trigger selection.

For the signal region (SR) selection, events satisfying the trigger requirements are required
to have at least one muon removal $\tauhad$ object which satisfies $\pt > 15~\GeV$, one or three reconstructed charged-particle tracks with charge summing to one, and a TauID jet RNN score $> 0.1$,
excluding $1.37 < |\eta| < 1.52$ and $|\eta| > 2.5$. This selection
implies that at least one reconstructed muon satisfying the `Medium' working point is inside the cone of each selected
$\tauhadmurm$ candidate. Additionally, the muon must have $\pt > 10~\GeV$ and opposite charge to the $\tauhadmurm$ candidate.
In this study, the overlap removal between muons and $\tauhad$ candidates is turned off as
the default overlap removal algorithm as described in Ref.~\cite{HIGG-2019-09} would remove
some of the $\tauhadmurm$ candidates even if the muon removal method is successful.
To suppress background from events containing heavy-flavour jets, events are rejected if they
contain any jet that satisfies the DL1d based $b$-tagging algorithm at an 85\% efficiency working point~\cite{FTAG-2019-07}.
The signed $\deltaphi$ between the muon and $\vecptmiss$ ($\deltamumet$) is required to be $-0.1 < \deltamumet < 0.4$.
The sign of $\deltamumet$ is determined by the direction of the $\tauhadmurm$ candidate.
If the $\vecptmiss$ is inside the opening angle of the muon and the $\tauhad$ candidate or if the $\vecptmiss$ is outside
the opening angle but closer to the muon, then the sign is positive; otherwise, it is negative.
Since the focus of the analysis is the boosted $\Zttmuhad$ process, a loose requirement $\mcol > 40$~\GeV,
and a requirement $\ptcol > 250$~\GeV are applied.
A validation region (VR) is defined with the same selection requirements as the SR, except that the
muon and $\tauhadmurm$ candidate are required to have the same charge, and there is no requirement on $\deltamumet$ or
the number of jets satisfying the $b$-tagging criteria.
The VR is designed with enough statistical
power for testing the background modelling. These tests confirm that multi-jets
events contribute negligibly to the background in both the SR and VR, making
it an unlikely source of significant background mismodelling.
Table~\ref{tab:evt_sel} summarises the event selection requirements for the SR and VR respectively.
\begin{DIFnomarkup}\begin{table}[b]
\caption{Summary of event selection requirements.}
\label{tab:evt_sel}
\centering
\scriptsize
\begin{tabular}{L{2cm}C{5.5cm}C{5.5cm}}
\toprule[1pt]
Object                                      & Signal region                                                 & Validation region                                      \\ \midrule[1pt]
\multirow{4}{*}{$\tauhadmurm$ candidate}    & $\pt > 15$~\GeV                                               & $\pt > 15$~\GeV                                        \\
& TauID jet RNN score $>$ 0.1                                   & TauID jet RNN score $>$ 0.1                           \\
& $|\eta| < 1.37$ or $1.52 < |\eta| < 2.5$                      & $|\eta| < 1.37$ or $1.52 < |\eta| < 2.5$              \\
& 1 or 3 charged-particle tracks                                & 1 or 3 charged-particle tracks                                 \\ \midrule
\multirow{2}{*}{Muon}                       & `Medium' working point                                        & `Medium' working point                                         \\
& $\pt > 10~\GeV$                                               & $\pt > 10~\GeV$                                       \\ \midrule
\multirow{5}{*}{$\tmth$ pair}               & no $b$-tagged jet at 85\% efficiency working point            & \textbf{---}                                            \\
& $-0.1 < \deltamumet < 0.4$                                    & \textbf{---}                                            \\
& $\mcol > 40~\GeV$                                             & $\mcol > 40~\GeV$                                     \\
& $\ptcol > 250~\GeV$                                           & $\ptcol > 250~\GeV$                                   \\
& \textbf{opposite charge} muon and $\tauhadmurm$ candidates    & \textbf{same charge} muon and $\tauhadmurm$ candidates   \\ \bottomrule[1pt]
\end{tabular}
\end{table}\end{DIFnomarkup}

Figures~\ref{fig:RNN_score}~and~\ref{fig:pt_col} illustrate the distributions of the TauID jet RNN score and $\ptcol$ in the SR and VR\@.
As shown in both figures, the data and MC predictions
agree well in both the SR and VR\@.
Due to the $\mcol > 40$~\GeV requirement, the kinematics of the selection produce a peak in the $\ptcol$ distribution around 500~\GeV.
This highlights the fact that the $\tauhadmurm$ reconstruction picks up $\tmth$ pairs only with a sufficiently high boost.
\begin{DIFnomarkup}\begin{figure}[tbp]
\centering
\subfloat[]{
\includegraphics[width=0.40\textwidth]{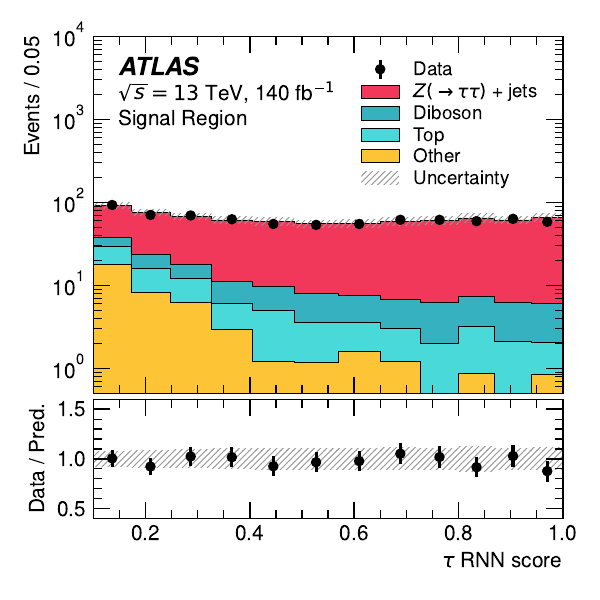}
\label{fig:SR_RNN_score}
}
\subfloat[]{
\includegraphics[width=0.40\textwidth]{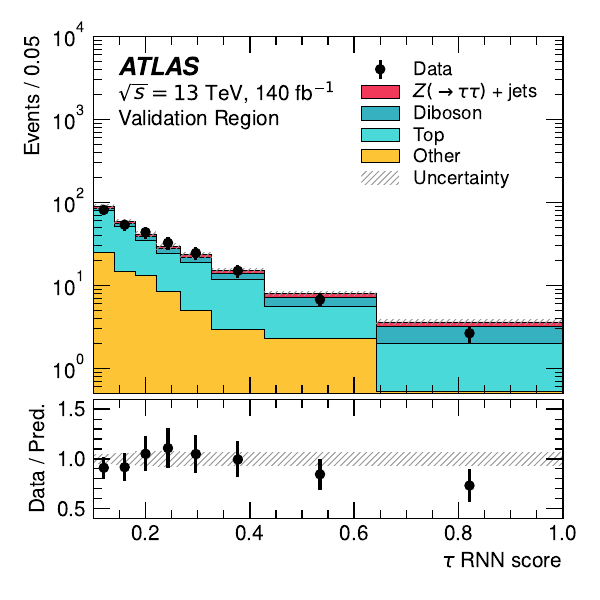}
\label{fig:CR_RNN_score}
}
\caption{
The distribution of the TauID jet RNN score for $\tauhadmurm$ (a) in the SR and
(b) in the VR.
`$\zttplusjets$' represents the contributions from the signal process.
`Top' represents the predicted contributions from the $\ttbar$, single-top-quark, and $tW$ processes.
`Diboson' indicates the contributions from $WW$, $WZ$, and $ZZ$ processes.
`Other' includes the contributions from the $Z(\rightarrow\ell\ell)$+jets, $W$+jets, and Higgs boson processes.
The uncertainties shown include both statistical and systematic sources.
}
\label{fig:RNN_score}
\end{figure}\end{DIFnomarkup}

\begin{DIFnomarkup}\begin{figure}[tbp]
\centering
\subfloat[]{
\includegraphics[width=0.40\textwidth]{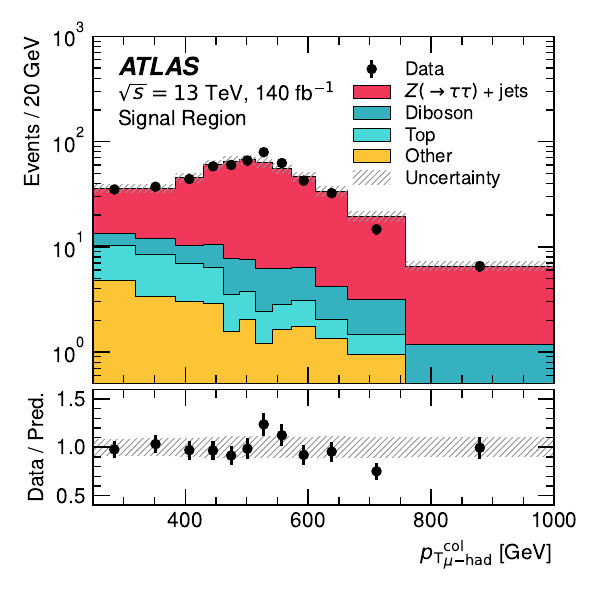}
\label{fig:SR_pt_col}
}
\subfloat[]{
\includegraphics[width=0.40\textwidth]{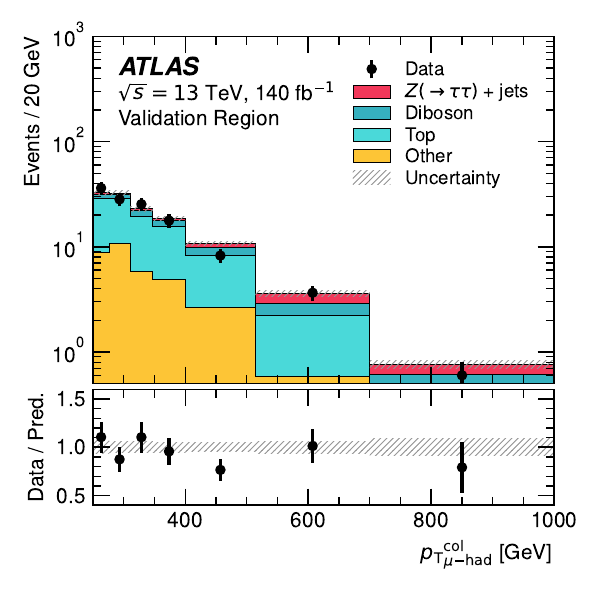}
\label{fig:CR_pt_col}
}
\caption{
The distribution of the $\ptcol$ (a) in the SR, and
(b) in the VR.
`$\zttplusjets$' represents the contributions from the signal process.
`Diboson' indicates the contributions from $WW$, $WZ$, and $ZZ$ processes.
`Top' represents the predicted contributions from the $\ttbar$, single-top-quark, and $tW$ processes.
`Other' includes the contributions from the $Z(\rightarrow\ell\ell)$+jets, $W$+jets, and Higgs boson processes.
The uncertainties shown include both statistical and systematic sources.
}
\label{fig:pt_col}
\end{figure}\end{DIFnomarkup}

\subsection{Systematic uncertainties} \label{sec:sys_uncertainties}
The dominant source of systematic uncertainty in the comparison between the observed and expected yields in the SR is the modelling of the cross-section
for $Z$-boson production. As discussed in Section~\ref{sec:DataAndMCs}, a +10\%
correction is applied to the predicted values for $\zttplusjets$ events,
with the full size of this correction quoted as a systematic uncertainty.
The most significant sources of experimental systematic uncertainties are related to TauID and $\tau$-lepton energy scale~(4\%)~\cite{ATLAS-CONF-2017-029, ATL-PHYS-PUB-2019-033},
jet energy scale and resolution~(2\%)~\cite{JETM-2018-05}, \MET~(2\%)~\cite{JETM-2020-03}, and luminosity~(0.83\%)~\cite{DAPR-2021-01}.
\subsection{Results} \label{sec:results}
To understand the performance improvement relative to
the standard ATLAS $\tauhad$ reconstruction and identification,
Figure~\ref{fig:SRcmp} shows multiple comparisons of
the data and MC predictions for \mcol\ distributions corresponding to various signal selections;
Figure~\ref{fig:SR_m} corresponds to the SR selections defined in Section~\ref{sec:ZttSelection};
Figure~\ref{fig:SR_std_m} corresponds to the \SRstd, which uses the standard ATLAS $\tauhad$ candidates,
without the muon removal, but otherwise corresponds to the same event selection as the SR;
Figure~\ref{fig:SR_tight_m} corresponds to the \SRtight, which
imposes an additional `Tight' RNN TauID requirement on the $\tauhadmurm$ candidates, but otherwise corresponds
to the SR selection;
and Figure~\ref{fig:SR_tight_std_m} corresponds to the \SRtightstd, which
imposes an additional `Tight' RNN TauID requirement and uses the standard $\tauhad$ candidates, without muon removal.

The collinear mass reconstruction of the di-$\tau$ system is quite
effective as it clearly shows the peak corresponding to the $Z$ boson mass in Figure~\ref{fig:SR_m}.
The shape of the $\mcol$ distribution for signal
Drell--Yan events is very different from that seen
in inclusive production~\cite{STDM-2018-17}, with a much larger fraction of the signal events in the
region $40~\GeV < \mcol < 70$~\GeV compared to that at the $Z$ boson peak.
This pattern arises due to the steep fall in the $\mcol$ distribution
for Drell--Yan production, along with the narrowing opening angle of
boosted $\tmth$ systems as $\mcol$ decreases.

\begin{DIFnomarkup}\begin{figure}[tbp]
\centering
\subfloat[]{
\includegraphics[width=0.40\textwidth]{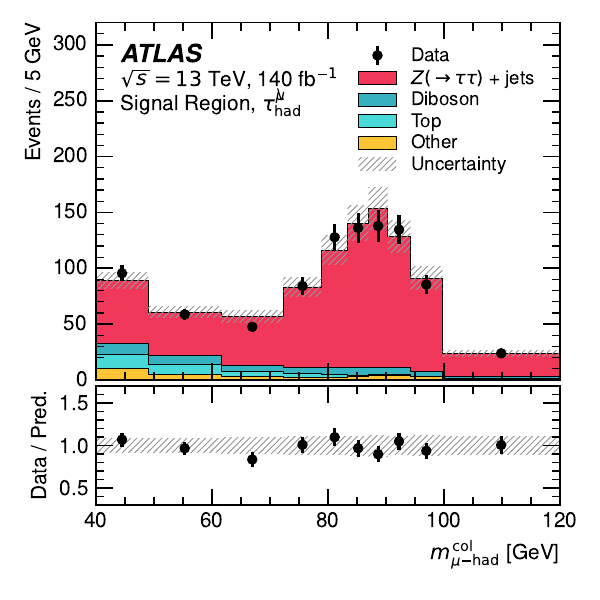}
\label{fig:SR_m}
}
\subfloat[]{
\includegraphics[width=0.40\textwidth]{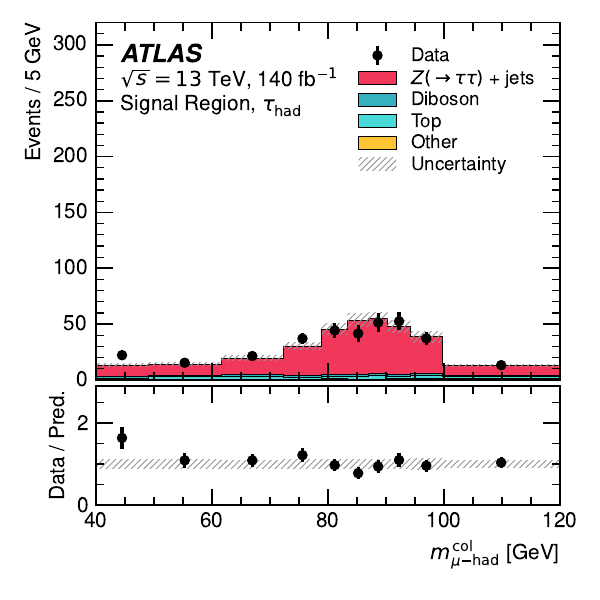}
\label{fig:SR_std_m}
}
\\
\subfloat[]{
\includegraphics[width=0.40\textwidth]{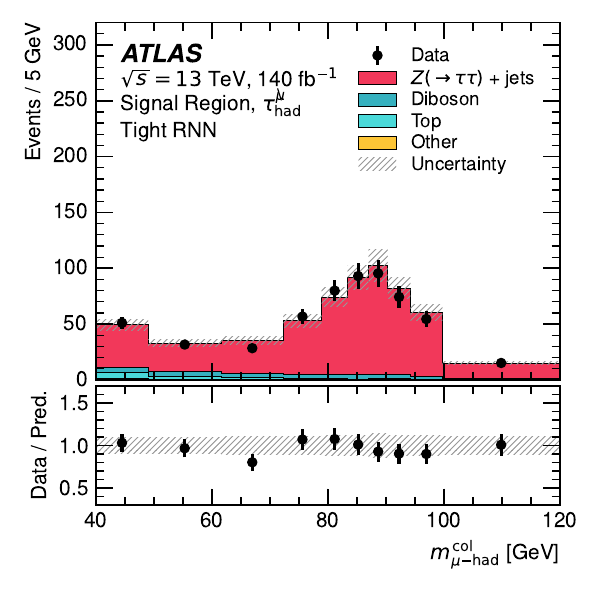}
\label{fig:SR_tight_m}
}
\subfloat[]{
\includegraphics[width=0.40\textwidth]{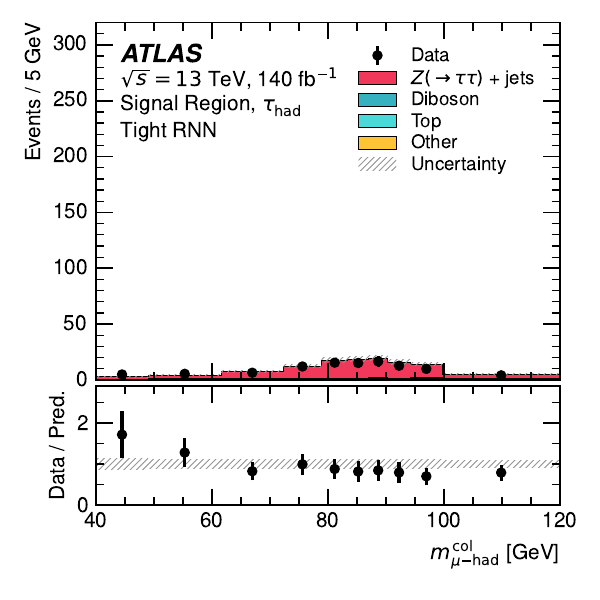}
\label{fig:SR_tight_std_m}
}
\caption{
The distributions of $\mcol$ corresponding to the various
signal region selection criteria defined in the text:
(a)~SR,
(b)~\SRstd,
(c)~\SRtight, and
(d)~\SRtightstd.
`$\zttplusjets$' represents the contributions from the signal process.
`Diboson' indicates the contributions from $WW$, $WZ$, and $ZZ$ processes.
`Top' represents the predicted contributions from the $\ttbar$, single-top-quark, and $tW$ processes.
`Other' includes the contributions from the $Z(\rightarrow\ell\ell)$+jets, $W$+jets, and Higgs boson processes.
The uncertainties shown include both statistical and systematic sources.
}
\label{fig:SRcmp}
\end{figure}\end{DIFnomarkup}
Table~\ref{tab:n_evt} shows the event yields in the data, as well as the predicted contributions from signal and background processes, corresponding to the various signal and validation selection criteria defined previously.
Compared with the standard TauID, the $\tauhadmurm$ method results in around
three times more signal events in the SR, accompanied by an increase in the number of background events.
In the \SRtight\, the number of signal events
is about five times higher than when using the standard $\tauhad$ candidates, with a corresponding rise in background events as well.
\begin{DIFnomarkup}\begin{table}[t!]
\caption{
Event yields for different event selection criteria, as defined in the text.
The uncertainties quoted are statistical only.
}
\label{tab:n_evt}
\centering
\scriptsize
\begin{tabular}{L{2cm}|R{1.2cm}R{1.2cm}|R{1.2cm}R{1.2cm}|R{1.2cm}}
\toprule[1pt]
& \SR                                                   & \SRstd                        & \SRtight                       & \SRtightstd                      & \VR                                       \\ \midrule[1pt]
$\Ztautau$              & $\textcolor{white}{0}945\pm \textcolor{white}{0}8$    & $\textcolor{white}{}325\pm 4$ & $\textcolor{white}{}633 \pm 7$ & $\textcolor{white}{}117 \pm 2$   & $\textcolor{white}{0}20 \pm 1$            \\
Diboson                 & $\textcolor{white}{00}91\pm \textcolor{white}{0}1$    & $\textcolor{white}{0}25\pm 1$ & $\textcolor{white}{0}50 \pm 1$ & $\textcolor{white}{00}9 \pm 0$   & $\textcolor{white}{0}38 \pm 1$            \\
Top                     & $\textcolor{white}{00}68\pm \textcolor{white}{0}3$    & $\textcolor{white}{0}53\pm 3$ & $\textcolor{white}{0}26 \pm 2$ & $\textcolor{white}{0}21 \pm 2$   & $\textcolor{white}{}164 \pm 5$            \\
Other                   & $\textcolor{white}{00}65\pm \textcolor{white}{0}4$    & $\textcolor{white}{0}17\pm 2$ & $\textcolor{white}{0}12 \pm 2$ & $\textcolor{white}{00}5 \pm 1$   & $\textcolor{white}{0}76 \pm 4$            \\ \midrule
Total predicted         & $\textcolor{white}{}1168\pm \textcolor{white}{}10$    & $\textcolor{white}{}420\pm 5$ & $\textcolor{white}{}721 \pm 8$ & $\textcolor{white}{}153 \pm 3$   & $\textcolor{white}{}297 \pm 7$            \\
Data                    & $\textcolor{white}{}1143$                             & $\textcolor{white}{}435$      & $\textcolor{white}{}698$       & $\textcolor{white}{}133$         & $\textcolor{white}{}281$                  \\
\bottomrule[1pt]
\end{tabular}
\end{table}\end{DIFnomarkup}
Subtracting the expected background yield in the SR of $223\pm 5$ events gives a
measured yield for $\Zttmuhad$ of $920 \pm 34$ events.
This can be compared with the expected yield for \Zttmuhad\ of $945 \pm 8~\text{(stat.)}\pm 114~\text{(syst.)}$ events.
Adding all sources of statistical and systematic uncertainty results in a total uncertainty of 12\%.
The same calculations are performed in the $\SRtight$.
The ratio of data to prediction is $0.97 \pm 0.12$ in the SR,
and $0.96 \pm 0.12$ in the $\SRtight$.

\FloatBarrier

\section{Conclusions} \label{sec:conclusion}
In this paper, the reconstruction of a highly boosted pair of $\tau$-leptons is considered,
in which one $\tau$-lepton decays into a muon plus two neutrinos
and the other $\tau$-lepton decays hadronically,
with the visible decay products overlapping.
In such cases, the standard ATLAS reconstruction and
identification algorithms for hadronically decaying $\tau$-leptons fail due
to the presence of the nearby muon.

The development of a muon removal procedure for $\tauhad$ candidates (denoted $\tauhadmurm$) using samples of high-mass $\GHHFourtau$
events is described. The $\tauhadmurm$ method recovers the \tauhad\ reconstruction and identification
efficiencies to the expected level for isolated \tauhad\ decays at all $\tau$-lepton identification working
points. The measurement precision for the charge and kinematic properties of the visible \tauhad\
system is similarly recovered. The background rejection power is studied using heavy-flavour jets
from top-quark decays and shows no degradation
relative to the standard algorithm for $\tauhad$ identification.
The $\tauhadmurm$ method is validated by selecting
a sample of highly boosted $\Zttmuhad$ final states
using a data set of proton--proton collisions
at $\sqrt{s}=13~\TeV$ collected with the ATLAS detector
at the LHC, and corresponding to an integrated luminosity of 140~\ifb.
Good agreement is found between data and simulation
in both the $\Zttmuhad$ signal region and a background-rich validation region.
The ratio of the observed and predicted
event yields in the signal region is
found to be consistent with unity
within an uncertainty of 12\%.

The results presented in this paper demonstrate the effectiveness of the $\tauhadmurm$
method in enhancing the signal sensitivity of the boosted $\tmth$ channel. The
observed good agreement between
data and simulation in the $\Zttmuhad$ signal region
reaffirms the robustness of the newly developed reconstruction method.
\section*{Acknowledgements}

%

%
%

%
%

We thank CERN for the very successful operation of the LHC and its injectors, as well as the support staff at
CERN and at our institutions worldwide without whom ATLAS could not be operated efficiently.

The crucial computing support from all WLCG partners is acknowledged gratefully, in particular from CERN, the ATLAS Tier-1 facilities at TRIUMF/SFU (Canada), NDGF (Denmark, Norway, Sweden), CC-IN2P3 (France), KIT/GridKA (Germany), INFN-CNAF (Italy), NL-T1 (Netherlands), PIC (Spain), RAL (UK) and BNL (USA), the Tier-2 facilities worldwide and large non-WLCG resource providers. Major contributors of computing resources are listed in Ref.~\cite{ATL-SOFT-PUB-2025-001}.

We gratefully acknowledge the support of ANPCyT, Argentina; YerPhI, Armenia; ARC, Australia; BMWFW and FWF, Austria; ANAS, Azerbaijan; CNPq and FAPESP, Brazil; NSERC, NRC and CFI, Canada; CERN; ANID, Chile; CAS, MOST and NSFC, China; Minciencias, Colombia; MEYS CR, Czech Republic; DNRF and DNSRC, Denmark; IN2P3-CNRS and CEA-DRF/IRFU, France; SRNSFG, Georgia; BMBF, HGF and MPG, Germany; GSRI, Greece; RGC and Hong Kong SAR, China; ICHEP and Academy of Sciences and Humanities, Israel; INFN, Italy; MEXT and JSPS, Japan; CNRST, Morocco; NWO, Netherlands; RCN, Norway; MNiSW, Poland; FCT, Portugal; MNE/IFA, Romania; MSTDI, Serbia; MSSR, Slovakia; ARIS and MVZI, Slovenia; DSI/NRF, South Africa; MICIU/AEI, Spain; SRC and Wallenberg Foundation, Sweden; SERI, SNSF and Cantons of Bern and Geneva, Switzerland; NSTC, Taipei; TENMAK, T\"urkiye; STFC/UKRI, United Kingdom; DOE and NSF, United States of America.

Individual groups and members have received support from BCKDF, CANARIE, CRC and DRAC, Canada; CERN-CZ, FORTE and PRIMUS, Czech Republic; COST, ERC, ERDF, Horizon 2020, ICSC-NextGenerationEU and Marie Sk{\l}odowska-Curie Actions, European Union; Investissements d'Avenir Labex, Investissements d'Avenir Idex and ANR, France; DFG and AvH Foundation, Germany; Herakleitos, Thales and Aristeia programmes co-financed by EU-ESF and the Greek NSRF, Greece; BSF-NSF and MINERVA, Israel; NCN and NAWA, Poland; La Caixa Banking Foundation, CERCA Programme Generalitat de Catalunya and PROMETEO and GenT Programmes Generalitat Valenciana, Spain; G\"{o}ran Gustafssons Stiftelse, Sweden; The Royal Society and Leverhulme Trust, United Kingdom.

In addition, individual members wish to acknowledge support from Armenia: Yerevan Physics Institute (FAPERJ); CERN: European Organization for Nuclear Research (CERN DOCT); Chile: Agencia Nacional de Investigaci\'on y Desarrollo (FONDECYT 1230812, FONDECYT 1230987, FONDECYT 1240864); China: Chinese Ministry of Science and Technology (MOST-2023YFA1605700, MOST-2023YFA1609300), National Natural Science Foundation of China (NSFC - 12175119, NSFC 12275265, NSFC-12075060); Czech Republic: Czech Science Foundation (GACR - 24-11373S), Ministry of Education Youth and Sports (FORTE CZ.02.01.01/00/22\_008/0004632), PRIMUS Research Programme (PRIMUS/21/SCI/017); EU: H2020 European Research Council (ERC - 101002463); European Union: European Research Council (ERC - 948254, ERC 101089007, ERC, BARD, 101116429), European Union, Future Artificial Intelligence Research (FAIR-NextGenerationEU PE00000013), Italian Center for High Performance Computing, Big Data and Quantum Computing (ICSC, NextGenerationEU); France: Agence Nationale de la Recherche (ANR-20-CE31-0013, ANR-21-CE31-0013, ANR-21-CE31-0022, ANR-22-EDIR-0002); Germany: Baden-Württemberg Stiftung (BW Stiftung-Postdoc Eliteprogramme), Deutsche Forschungsgemeinschaft (DFG - 469666862, DFG - CR 312/5-2); Italy: Istituto Nazionale di Fisica Nucleare (ICSC, NextGenerationEU), Ministero dell'Università e della Ricerca (PRIN - 20223N7F8K - PNRR M4.C2.1.1); Japan: Japan Society for the Promotion of Science (JSPS KAKENHI JP22H01227, JSPS KAKENHI JP22H04944, JSPS KAKENHI JP22KK0227, JSPS KAKENHI JP23KK0245); Netherlands: Netherlands Organisation for Scientific Research (NWO Veni 2020 - VI.Veni.202.179); Norway: Research Council of Norway (RCN-314472); Poland: Ministry of Science and Higher Education (IDUB AGH, POB8, D4 no 9722), Polish National Agency for Academic Exchange (PPN/PPO/2020/1/00002/U/00001), Polish National Science Centre (NCN 2021/42/E/ST2/00350, NCN OPUS 2023/51/B/ST2/02507, NCN OPUS nr 2022/47/B/ST2/03059, NCN UMO-2019/34/E/ST2/00393, NCN \& H2020 MSCA 945339, UMO-2020/37/B/ST2/01043, UMO-2021/40/C/ST2/00187, UMO-2022/47/O/ST2/00148, UMO-2023/49/B/ST2/04085, UMO-2023/51/B/ST2/00920, UMO-2024/53/N/ST2/00869); Slovenia: Slovenian Research Agency (ARIS grant J1-3010); Spain: Generalitat Valenciana (Artemisa, FEDER, IDIFEDER/2018/048), Ministry of Science and Innovation (MCIN \& NextGenEU PCI2022-135018-2, MICIN \& FEDER PID2021-125273NB, RYC2019-028510-I, RYC2020-030254-I, RYC2021-031273-I, RYC2022-038164-I); Sweden: Carl Trygger Foundation (Carl Trygger Foundation CTS 22:2312), Swedish Research Council (Swedish Research Council 2023-04654, VR 2018-00482, VR 2022-03845, VR 2022-04683, VR 2023-03403, VR grant 2021-03651), Knut and Alice Wallenberg Foundation (KAW 2018.0458, KAW 2019.0447, KAW 2022.0358); Switzerland: Swiss National Science Foundation (SNSF - PCEFP2\_194658); United Kingdom: Leverhulme Trust (Leverhulme Trust RPG-2020-004), Royal Society (NIF-R1-231091); United States of America: U.S. Department of Energy (ECA DE-AC02-76SF00515), Neubauer Family Foundation.

%
%


\clearpage
\printbibliography
\clearpage
 
\begin{flushleft}
\hypersetup{urlcolor=black}
{\Large The ATLAS Collaboration}

\bigskip

\AtlasOrcid[0000-0002-6665-4934]{G.~Aad}$^\textrm{\scriptsize 104}$,
\AtlasOrcid[0000-0001-7616-1554]{E.~Aakvaag}$^\textrm{\scriptsize 17}$,
\AtlasOrcid[0000-0002-5888-2734]{B.~Abbott}$^\textrm{\scriptsize 123}$,
\AtlasOrcid[0000-0002-0287-5869]{S.~Abdelhameed}$^\textrm{\scriptsize 119a}$,
\AtlasOrcid[0000-0002-1002-1652]{K.~Abeling}$^\textrm{\scriptsize 56}$,
\AtlasOrcid[0000-0001-5763-2760]{N.J.~Abicht}$^\textrm{\scriptsize 50}$,
\AtlasOrcid[0000-0002-8496-9294]{S.H.~Abidi}$^\textrm{\scriptsize 30}$,
\AtlasOrcid[0009-0003-6578-220X]{M.~Aboelela}$^\textrm{\scriptsize 46}$,
\AtlasOrcid[0000-0002-9987-2292]{A.~Aboulhorma}$^\textrm{\scriptsize 36e}$,
\AtlasOrcid[0000-0001-5329-6640]{H.~Abramowicz}$^\textrm{\scriptsize 155}$
% ...
\bigskip
\\

% ...
$^{17}$Department for Physics and Technology, University of Bergen, Bergen; Norway.\\
% ...
$^{30}$Physics Department, Brookhaven National Laboratory, Upton NY; United States of America.\\
% ...
$^{36}$$^{(a)}$Facult\'e des Sciences Ain Chock, Universit\'e Hassan II de Casablanca;$^{(b)}$Facult\'{e} des Sciences, Universit\'{e} Ibn-Tofail, K\'{e}nitra;$^{(c)}$Facult\'e des Sciences Semlalia, Universit\'e Cadi Ayyad, LPHEA-Marrakech;$^{(d)}$LPMR, Facult\'e des Sciences, Universit\'e Mohamed Premier, Oujda;$^{(e)}$Facult\'e des sciences, Universit\'e Mohammed V, Rabat;$^{(f)}$Institute of Applied Physics, Mohammed VI Polytechnic University, Ben Guerir; Morocco.\\
% ...
$^{46}$Physics Department, Southern Methodist University, Dallas TX; United States of America.\\
% ...
$^{50}$Fakult\"{a}t Physik , Technische Universit{\"a}t Dortmund, Dortmund; Germany.\\
% ...
$^{56}$II. Physikalisches Institut, Georg-August-Universit\"{a}t G\"ottingen, G\"ottingen; Germany.\\
% ...
$^{104}$CPPM, Aix-Marseille Universit\'e, CNRS/IN2P3, Marseille; France.\\
% ...
$^{119}$$^{(a)}$New York University Abu Dhabi, Abu Dhabi;$^{(b)}$United Arab Emirates University, Al Ain; United Arab Emirates.\\
% ...
$^{123}$Homer L. Dodge Department of Physics and Astronomy, University of Oklahoma, Norman OK; United States of America.\\
% ...
$^{155}$Raymond and Beverly Sackler School of Physics and Astronomy, Tel Aviv University, Tel Aviv; Israel.\\
% ...
$^{a}$ Also Affiliated with an institute covered by a cooperation agreement with CERN.\\
$^{b}$ Also at An-Najah National University, Nablus; Palestine.\\
$^{c}$ Also at Borough of Manhattan Community College, City University of New York, New York NY; United States of America.\\
$^{d}$ Also at Center for High Energy Physics, Peking University; China.\\
$^{e}$ Also at Center for Interdisciplinary Research and Innovation (CIRI-AUTH), Thessaloniki; Greece.\\
$^{f}$ Also at CERN, Geneva; Switzerland.\\
% ...

\end{flushleft}


%

%
%

\end{document}